\documentclass[10pt,journal,compsoc]{IEEEtran}
\ifCLASSOPTIONcompsoc
  \usepackage[nocompress]{cite}
\else
  \usepackage{cite}
\fi
\ifCLASSINFOpdf
\else
\fi
\usepackage{url}

\usepackage{soul}
\usepackage{tabularx}
\usepackage{xcolor}
\usepackage{caption}
\usepackage{threeparttable}
\usepackage{graphicx}
\usepackage{booktabs}
\usepackage{colortbl}
\usepackage{comment}
\usepackage{rotating}
\usepackage{pdflscape}
\usepackage{fancyhdr} 

\fancypagestyle{mylandscape}{
\fancyhf{} 
\fancyfoot{
\makebox[\textwidth][c]{
  \rlap{\hspace{.75cm}
    \smash{
      \raisebox{4.87in}{
        \rotatebox{90}{\thepage}}}}}}
}

\renewcommand\hl[1]{#1}

\begin{document}
%
\title{Sixteen Years of Phishing User Studies: What Have We Learned?}
%
%
%
%

\author{Shahryar~Baki,
        Rakesh~Verma,~\IEEEmembership{Member,~IEEE,}
\thanks{This work has been submitted to the IEEE for possible publication. Copyright may be transferred without notice, after which this version may no longer be accessible.\\S. Baki and R. Verma were with the Computer Science Department, University of Houston, Houston,
TX, 77004 USA e-mail: {sbaki2,rmverma2}@uh.edu).}}

%
%

\markboth{Journal of \LaTeX\ Class Files,~Vol.~14, No.~8, August~2015}%
{Baki \MakeLowercase{\textit{et al.}}: Sixteen Years of Phishing User Studies: What Have We Learned?}
%



\IEEEtitleabstractindextext{%
\begin{abstract}
Several previous studies  have investigated  user 
susceptibility to phishing attacks. A thorough meta-analysis or systematic review is required to gain a better understanding of these findings \hl{and to assess the strength of evidence for phishing susceptibility of a subpopulation, e.g., older users. We aim to determine whether an effect exists; another aim is to determine whether the effect is positive or negative and to obtain a single summary estimate of the effect.}

\noindent\textbf{OBJECTIVES:} We systematically review the results of previous user studies on phishing susceptibility and conduct a meta-analysis. 

\noindent\textbf{METHOD:} 
We searched four online databases 
for English studies on phishing. 
\hl{We included all user studies in phishing} detection and prevention, whether \hl{they proposed new training techniques} or \hl{analyzed users'} vulnerability. 

\noindent\textbf{FINDINGS:} A careful analysis reveals some discrepancies between the findings. More than half of the studies that analyzed the effect of \textit{age} reported no statistically significant relationship between age and users' performance. \hl{Some studies reported older people performed better while some reported the opposite}.
A similar finding holds for the gender difference. 
The meta-analysis 
\hl{shows: 1) a significant relationship between participants' age and their susceptibility 2) females are more susceptible than males 3) users training significantly improves their detection ability.} 
\end{abstract}

\begin{IEEEkeywords}
phishing, spear phishing, email, website, systematic review, meta-analysis, survey, user survey, training, highlighting, study, \hl{Uniform Resource Locator (URL)}
\end{IEEEkeywords}}

\maketitle

\IEEEdisplaynontitleabstractindextext

%
\IEEEpeerreviewmaketitle

\IEEEraisesectionheading{\section{Introduction}\label{sec:introduction}}

\IEEEPARstart{T}{he} \hl{email from Charlie stated blandly, ``here is a free pdf of the popular book on the president,'' with the book's title in bold font. Alice had downloaded it, printed it out and had been engrossed in reading it. But a few hours later, her computer started having issues. Then all her files were encrypted and there was a demand on the screen for payment of 1.0 bitcoin (worth about
\$40,000 at the time) to get her data back. It was at this point that Alice  realized she'd not heard from Charlie for a while, that his email had seemed a bit strange, coming out of the blue like that. And, of course, she definitely should not have downloaded that pdf ...}

\hl{Alice may not be real, but there are many people like her who have fallen victim to email-based attacks such as this. These emails follow a pattern: they pretend to be from a person or organization that the recipient knows and trusts, they encourage people to click a link, or download a file, or visit a website and input their username and password, only for them to later discover that their identity or data has been compromised.}

Emails and messages carefully designed to steal the recipients sensitive information or to install malware are called phishing attacks. Although phishing has been around since at least 1995 \cite{das2019} \hl{(the term phishing was used in 1996 for the first time)}, it still continues to be successful and a weapon of choice for attackers. The reason for this success is that these attacks target the human recipient and do not need to penetrate any technical defenses deployed, e.g., firewalls or intrusion detectors. \hl{Although the growth of social networks attackers has increased the attack surface, email is still the most common vector used by attackers} \cite{das2019}. \hl{These emails themselves usually contain a link to a fake website (probably identical to the original website) to steal users' sensitive information. Phishing attacks have been successful against big companies like RSA.} \cite{diehl16}

Humans are known as the weakest link in defending against security attacks \cite{sasse2001transforming}, and phishing is no exception. Phishing training programs are mostly ``shooting in the dark,'' i.e., they have been mostly trying to increase people's knowledge and experience without paying much attention to what academic studies are telling them. 
Researchers have been studying the relation between users' demographics and their susceptibility to phishing attacks  \cite{lastdrager2017effective, dodge2006using}. Some of them found statistically significant relationships and some did not. There is also some evidence of the effect of technological and background knowledge on users' ability in distinguishing attacks from legitimate scenarios \cite{kearney2013phishing, blythe2011f}.  After reviewing these studies \cite{das2019}, we discovered that there are some discrepancies between their findings. Hence, a more systematic comparison is necessary.  \hl{Listed below are the questions we seek to answer} \hl{based on individual and external attributes}:\footnote{we use these variables since they have been studied in many \hl{(more than 150)} previous works \cite{das2019}}
\begin{enumerate}
    \item Age: Do older users behave differently from younger, i.e., are they significantly more or less susceptible to phishing attacks?
    \item Gender: Is there any difference between detection performance of male and female users?
    \item Technological Background: Does having more or less background knowledge affect phishing susceptibility?
    \item Training: Can training be relied on as a mechanism to improve users security awareness/knowledge?
    \item Warning/highlighting: Does showing a warning or highlighting parts of the email/website increase people's ability in distinguishing attacks from legitimate requests?
\end{enumerate}

We carefully review previous user studies on phishing 
and explore whether a meta-analysis of the research literature on phishing behaviors and susceptibility is possible. \hl{Several researchers studied different aspects of users and their susceptibility to phishing attacks. As we mention later in this paper, the findings of previous works sometimes contradict each other (partly due to the differences in their designs) and can be confusing. Meta-analysis is the best way to combine their findings to identify the pattern between the studies' findings and derive the pooled estimate of the truth }\cite{greenland2008meta}. Our findings indicate that the results presented in phishing user studies leave much to be desired. Many studies (45 out of 82) did not report enough data to calculate the effect size and confidence interval for generating the meta-analysis model.
Hence, we conduct a systematic review and use the subset of studies that reported enough information for meta-analysis. We hope that our paper will inspire more discussion about user studies on phishing. We also hope to convince researchers to report their results in a more scientific and comprehensive manner so that a meta-analysis can be conducted and provide convincing evidence on which to base training regimes. Our analysis also shows gaps in the literature where more research is needed.

\section{Methods}~\label{sec:method}
Here we describe the collection  procedure for research articles that conducted user studies on phishing detection/susceptibility. Then we elaborate on the criteria used for filtering the research articles. 
We follow the steps suggested in \cite{russo2007review} to conduct the meta-analysis.

\subsection{Database Search}
Four databases were used independently to collect research papers: DBLP, ACM Digital Library, IEEE Xplore, and ProQuest Dissertations/Theses Global. All searches were conducted in February 2020. First, we started with queries \textit{``phish URL, phish link, phish site, phish web, phish email/e-mail''} but then we found several papers that study phishing as part of malicious and malware-centric behavior. Hence, we added some more queries to our initial list (without quotes) - \textit{``malware URL, malicious URL, malware link, malicious link, malware site, malicious site, malware email/e-mail, malicious email/e-mail.''}
To search for relevant literature on spear phishing, we used the queries \textit{``spear phishing''} and \textit{``spearphishing.''} The queries \textit{``spear phish''} and \textit{``spearphish''}  did not yield any additional results.
Later, we realized that authors sometimes just use \textit{``phishing detection''} or \textit{``phishing attack,''} or some other variation. So we expanded our search using \textit{``phish''} and \textit{``phishing''} to the above databases.

\begin{figure}
    \centering
    \includegraphics[width=\columnwidth]{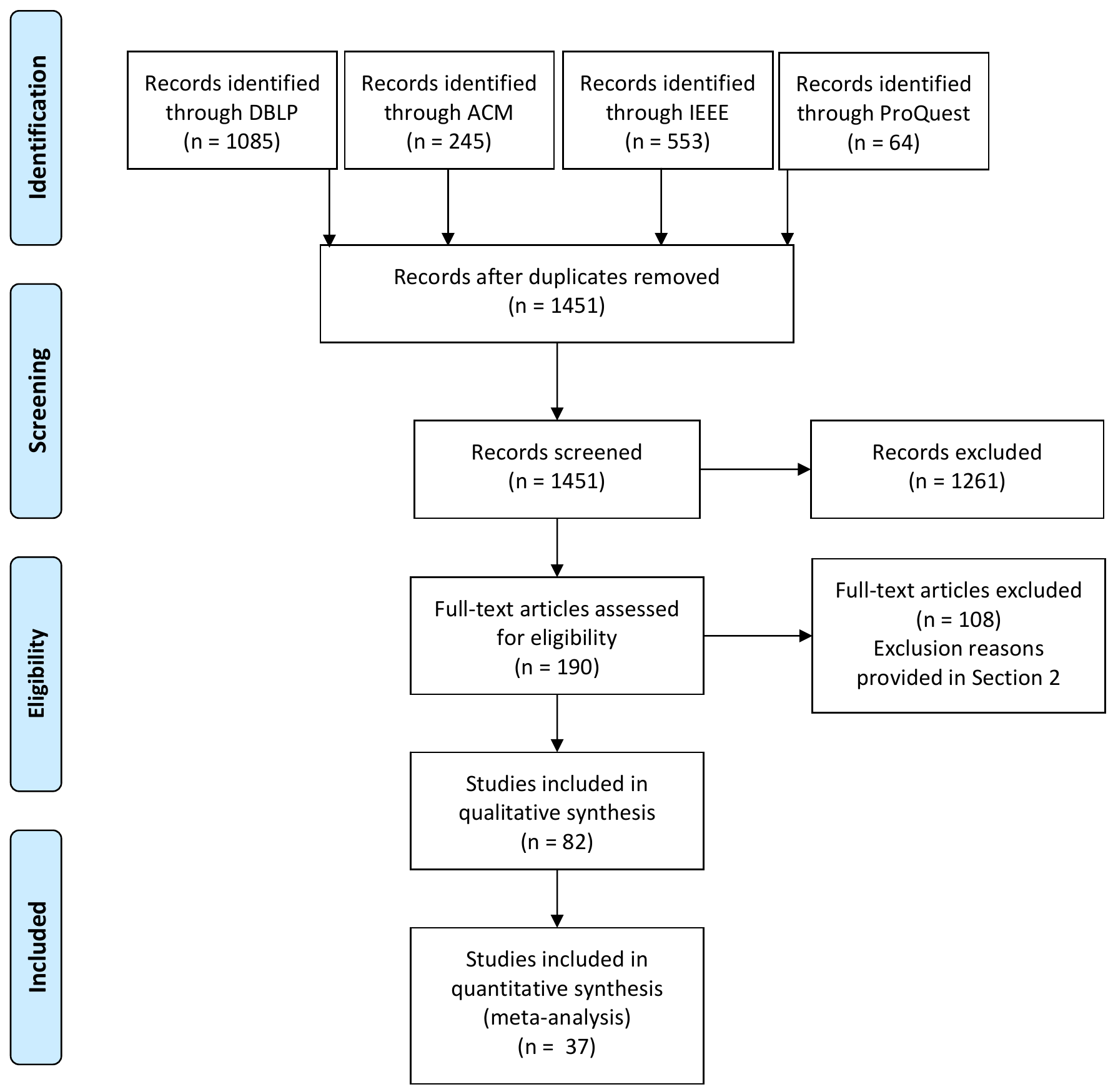}
    \caption{Study flow diagram.}
    \label{figure:PRISMA_flowchart}
\end{figure}

Database searches (no limit on publication year) returned 1,947 articles (Figure \ref{figure:PRISMA_flowchart}). \hl{The first phishing paper appeared in 2004.} Of the 1,947 total articles, 496 were identified as duplicates and removed. Most of the research on phishing is focused on proposing automated detection systems \cite{khonji2013phishing,das2019} and not on user studies. Thus, the remaining studies were screened based on the title and abstract to find those studies that did some kind of user experiment. Papers that conducted user experiments are kept for full-text assessment to extract their findings. If it was not clear based on the title and abstract that a research conducted a user study or not, we also kept it for full-text assessment. This screening step left 190 articles for full-text assessment.

\subsection{Inclusion and Exclusion Criteria}
This systematic review follows the guidelines defined in the Preferred Reporting Items for Systematic Reviews and Meta-Analyses \cite{knobloch2011preferred}. \hl{Although this guideline was initially designed for medical studies, it was soon adopted by researchers in other fields as well (e.g., image encryption and computer vision)} \cite{xu2020computer, SOKOUTI201814}. In particular, the PICOS framework is followed in defining the inclusion/exclusion criteria (Table \ref{table:criteria}). It is important to emphasize  that these criteria are defined first and then we start browsing the literature collected. For a study to be included, all the inclusion criteria must be met. Hence for exclusion, a single exclusion criterion is enough.

\begin{table*}
\caption{PICOS criteria for inclusion and exclusion of studies. All \hl{five} inclusion criteria must be met for a study to be included \hl{(for the rows with more than one criteria, separated by semicolon, meeting one of them is enough)}.}
\begin{tabular}{p{1.8cm} p{7.5cm} p{7.5cm}}
\toprule
\textbf{Component}             & \multicolumn{1}{c}{\textbf{Inclusion Criteria}}                 & \multicolumn{1}{c}{\textbf{Exclusion Criteria}}        \\[0.15cm] \hline
Population            & Any age and location                                   & Participants with any form of disability or disorder                                        \\[0.15cm]
Intervention & Some participants are shown warning/training; Older participants; males; participants with background knowledge  & Any other intervention, e.g. performance based on attack delivery medium                                          \\[0.15cm]
Comparators           & Some participants do not receive the proposed warning/training; Younger participants, females, or participants without background knowledge   & Lack of comparator (control) group     \\[0.15cm]
Outcome       & Ability in detecting phishing and legitimate instances & 
Reporting detection performance without statistical tests\\[0.15cm]
Study design          & Experimental         & Observational; Using knowledge approach instead of behavioral approach \\ \hline
\end{tabular}
\label{table:criteria}
\end{table*}

\subsubsection{Population}
We do not put any restrictions on the geographical location of participants, nor their language. There are \hl{only two studies} on the susceptibility of users with some kind of disability or disorder \cite{neupane2018social,blythe2011f}. Since their results are not directly comparable with other studies, and since we only found two such studies, we decided not to include them in this review \hl{(to control the heterogeneity of our models)}. 
In \cite{neupane2018social}, the experiment involved participants with social disorders, hence it is omitted. In \cite{blythe2011f}, researchers conducted two sets of experiments, one with a general population and one with blind people. We include the results of the general population experiment only.

\subsubsection{Intervention and Comparators}
\hl{Our main goal in this paper is to compare and combine findings} of previous works on users' susceptibility with respect to the following variables: age, gender, technological background, the effect of training, and the effect of warning. \hl{We call the first three, individual attributes and the other two external attributes since they are adopted by organizations or technologies.} We only include studies that compared users' phishing detection performance based on the aforementioned variables. For the comparator, we only consider studies that have at least one control group to compare with the intervention group. For example, in studying the effectiveness of training, we only include studies that compare a new training method with a control group (whether a group without any training or a group exposed to an existing training) as a baseline.

\subsubsection{Outcome}
Users' accuracy in detecting phishing and/or legitimate instances is the main variable to measure users' susceptibility to phishing attacks. Some works report the accuracy in detecting phishing and legitimate instances separately and some report them combined together. The design of the study can also affect the outcome variable. Some studies administer real world phishing attacks while some show the phishing/legitimate instances to participants and ask them to label the instances. So, the definition of the accuracy might be different between studies. We will discuss this later when we mention the bias in meta-analysis (Section \ref{sec:sub_gender_effect}).

\subsubsection{Study Design}
We only include the studies that exposed participants to some form of phishing/legitimate attacks, either through conducting a phishing attacks or role playing scenario where they ask participants to decide the legitimacy of a website or email. We exclude studies that use \hl{a knowledge based approach, i.e., studies that just ask participants about their behaviour in different scenarios instead of showing them  attacks}. \hl{Those studies are not directly comparable to the studies that show/execute the actual attacks as they measure users' metacongnition rather than their vulnerabilities.}



\subsection{Full-Text Assessment}
From the included studies, information was collected regarding the variables that have been studied. Studies were also evaluated for their risk of bias based on the methods of participant selection, environment (whether the study has been done in the lab or real-world), study design, and statistical analyses. 

Out of the 190 articles that qualified for full-text assessment, 108 articles did not meet our inclusion criteria and have been removed from the review. Below is the break down of the exclusion criteria applied in a sequential fashion. 
\begin{itemize}
    \item Thirty four studies did not study our goal variables.
    \item Thirty four studies used the knowledge based approach.
    \item Sixteen studies did not use any statistical methods to measure the significance of the reported differences. 
    \item Six studies just proposed methods without any evaluation (e.g. an idea of a training application). 
    \item There were five dissertations for which  authors published at least one paper in a journal or conference. So we only included the published version of these. We cross-checked their findings to make sure the dissertations do not have any extra details or results. 
    \item Three studies only did a qualitative study without showing any legitimate and phishing emails to participants. 
    \item There were also three papers for which  authors submitted a more detailed version of the same study (to either a journal or Arxiv) \cite{arachchilage2015can,alnajim2009evaluation, herzberg2004security}. So, we only considered the journal version of that article and ignored the conference version. 
    \item Four papers did not conduct user studies, e.g. authors in \cite{brooks2018} did a statistical analysis of the persuasive techniques used in phishing emails. 
    \item We could not find the full-text of a dissertation \cite{arora2019} (i.e., the author prohibited ProQuest from sharing it) and two published articles \cite{fatima2019persuasive, parsons2018}.\footnote{Two dissertations were not found in ProQuest database, but we received one of them \cite{martin2019} after contacting the authors of both.} 
   
\end{itemize}

\hl{As shown in Figure} \ref{figure:PRISMA_flowchart}, \hl{82 papers were included in the qualitative study and only 37 of them were included in the quantitative study, since not all of them reported their results in a way that allows meta analysis. We mention them later when we discuss the meta analysis results}.

\section{Results}
\begin{figure}
    \centering
    \includegraphics[width=0.5\columnwidth]{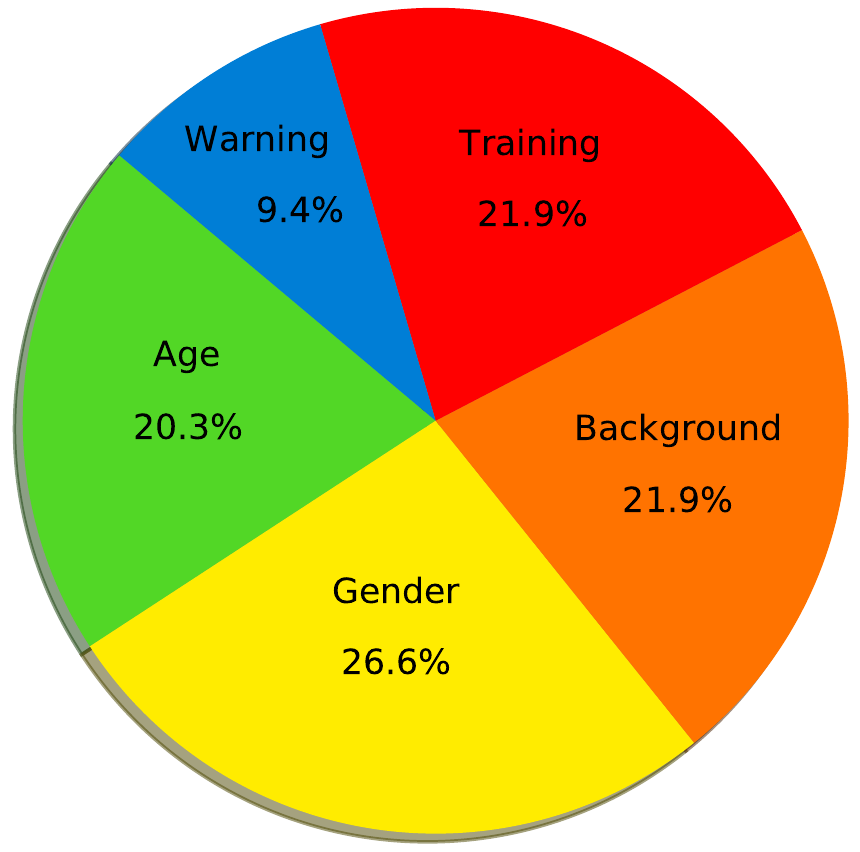}
    \caption{Percentage of reviewed papers that 
    analyzed each variable (some papers analyzed more than one variable, so the total number of papers is 128 instead of 82).}
    \label{figure:pie_stat}
\end{figure}

After filtering the papers in the literature based on the criteria mentioned above, we ended up with 76 papers and six dissertations (we call both of them ``paper'' in the rest of this article).
Figure \ref{figure:pie_stat} shows the percentage of papers that analyzed each variable (some papers analyzed more than one variable). 
The effect of \textit{age}, \textit{gender}, \textit{training}, and \textit{background knowledge} on phishing susceptibility have been studied by almost the same number of papers. The effect of \textit{warning} has been studied the least.

\begin{figure}
    \centering
    \includegraphics[width=\columnwidth]{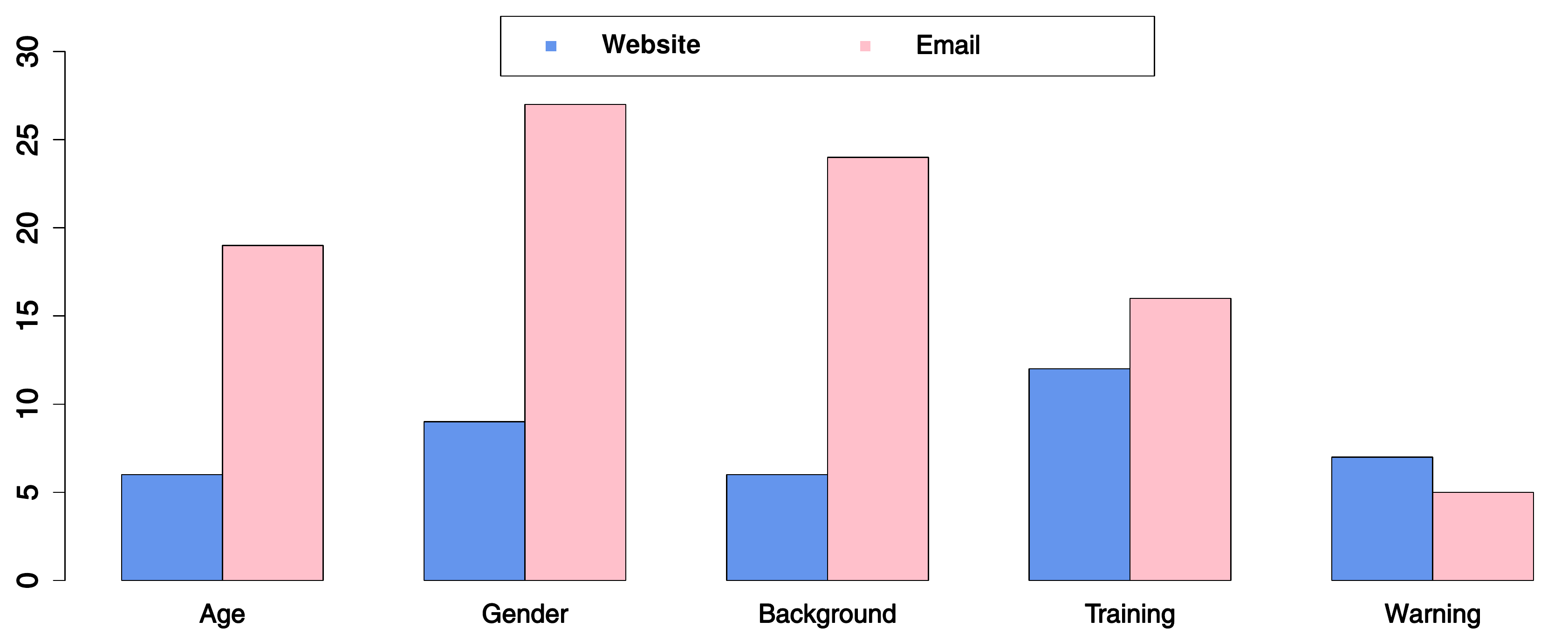}
    \caption{Number of studies that analyzed each variable categorized by attack vector.}~\label{figure:paper_stats}
\end{figure}

Figure \ref{figure:paper_stats} shows the number of papers that studied each variable in website and email contexts. As is apparent from the figure, more studies have been done on emails than on websites. \hl{Individual attributes (age, gender, and background knowledge)} are the most studied variables since most of the studies collect demographic data, even if their main goal is not these variables. For example, a study analyzing a new way of highlighting can concomitantly  evaluate the effect of age and gender without modifying the design. \hl{In the rest of this section, we first report our findings on individual attributes and then cover the external attributes.}

\begin{sidewaystable*}
\caption{Studies that compared users based on their age and gender (sorted based on their publication year). Green cells show significant findings. Brighter green shows opposite findings. Gender distribution column only shows male and female percentage (male:female), so if an study has a third category (e.g. others) their sum does not add up to 100 \cite{valipe2018, kumaraguru2010teaching}. SEM: Structural Equation Modeling, PLS: Partial Least-Squares. OLS: Ordinary Least-Squares.}
\resizebox{\textwidth}{!}{
\begin{tabular}{|c|r|c|c|c|c|c|c|}
\hline
\textbf{Paper} &  \textbf{Size} & \textbf{Vector} & \textbf{Age} & \textbf{Age Range} & \textbf{Gender} & \textbf{Gender dist.} & \textbf{Statistical Test} \\ \hline
Lin et. al. (2019) \cite{lin2019susceptibility} & 158 & Email & \begin{tabular}[l]{@{}l@{}}\textbf{Old}: most susceptible to reciprocation\\ \textbf{Young}: most susceptible to scarcity\end{tabular}&\begin{tabular}[l]{@{}l@{}}\textbf{Old}: $\bar{x}=61.7$, $\sigma=6.8$\\\textbf{Young}: $\bar{x}=21.7$, $\sigma=4.1$ \end{tabular} & \cellcolor{green!25}{\begin{tabular}[l]{@{}l@{}}\hl{Older women more susceptible} \end{tabular}} & \begin{tabular}[l]{@{}l@{}}\textbf{Old}: 57:43\\\textbf{Young}: 44:56 \end{tabular}   &Logistic Regression\\ \hline
Parsons et. al. (2019)\cite{parsons2019predicting} & 985 & Email & \cellcolor{green!25}{\begin{tabular}[l]{@{}c@{}}Susceptibility to phishing\\decreased with age \end{tabular}} & $\geq$18& \cellcolor{red!25}{No difference} & 47:53 & Hierarchical Regression\\ \hline
Musuva et. al. (2019) \cite{musuva2019new} & 192 & Email & \cellcolor{red!25}{No difference} & $\geq$18 & \cellcolor{red!25}{No difference} & 63:37& SEM\\ \hline
Taib et. al. (2019) \cite{taib2019social} & 56,365 & Email& \cellcolor{green!55}{\begin{tabular}[l]{@{}l@{}}Elders are more susceptible\end{tabular}}& $\geq$18 &\cellcolor{red!25}{\begin{tabular}[l]{@{}l@{}}No difference\end{tabular}} & 49.6:50.4 &Pearson’s $\chi^2$ \\ \hline
Orunsolu et. al. (2018) \cite{orunsolu2018users} & 427& Both & & $\leq$50 & \cellcolor{green!55}{\begin{tabular}[l]{@{}l@{}}Females performed better\end{tabular}} & 57:43 & Univariate Linear Regression\\ \hline
Diaz et. al. (2018) \cite{diaz2020phishing} & 1246 & Email& \cellcolor{green!25}{\begin{tabular}[l]{@{}l@{}}Elders are less susceptible\end{tabular}} & N/A & \cellcolor{red!25}{\begin{tabular}[l]{@{}l@{}}No difference\end{tabular}} & N/A & Pearson’s $\chi^2$\\ \hline
Valipe (2018) \cite{valipe2018} & 156 & Email & \cellcolor{green!25}{Elders are less susceptible} & 19-74 & \cellcolor{red!25}{No difference} & 59:36 & ANOVA\\ \hline
Lastdrager et. al. (2017) \cite{lastdrager2017effective}& 353 & \begin{tabular}[c]{@{}c@{}}Both\end{tabular}& \cellcolor{green!25}{Younger children are more vulnerable}& \begin{tabular}[c]{@{}c@{}}8-13 \\($\bar{x}=10.6, \sigma=1.0$) \end{tabular} &\cellcolor{red!25}{No difference} & 46:54 & \begin{tabular}[c]{@{}c@{}}Age: Linear Regression\\Gender: t-test\end{tabular}\\ \hline
Benenso et. al.(2017) \cite{benenson2017unpacking} & 1255 & \begin{tabular}[c]{@{}c@{}}Email vs.\\Facebook\end{tabular} & & N/A & \cellcolor{red!25}{No difference} & 27:73&  Pearson’s $\chi^2$\\ \hline
Goel et. al. \cite{goel2017got} & 7,225 & Email && N/A & \cellcolor{red!25}{No difference} & 51:49&  Pearson’s $\chi^2$ \\ \hline
Bullee et. al. (2017) \cite{bullee2017spear} & 593 & Email &\cellcolor{red!25}{No difference}& 22-76 & \cellcolor{red!25}{No difference} & 75.5:24.5 & Logistic Regression \\ \hline
Sun and Yeh (2017) \cite{sun2017effects} &80 &Email& & $\bar{x}=23.5$, $\sigma=3.2$ & \cellcolor{red!25}{No difference} & 54:46&  ANCOVA\\\hline
Butavicius et. al. (2017) \cite{butavicius2017understanding} &121& Email&\cellcolor{red!25}{No difference}& $\geq$18 &\cellcolor{red!25}{No difference}& 32:68 & Linear Regression   \\ \hline
Moody et. al. (2017) \cite{moody2017phish} & 595 & Email &\cellcolor{red!25}{No difference} & $\bar{x}=20.5$, $\sigma=2.4$ &\cellcolor{red!25}{No difference} & 56-44 & Logistic Regression \\ \hline
Canfield et. al. (2016) \cite{canfield2016quantifying} &152 & Email & \cellcolor{red!25}{No difference} & 19-59 ($\bar{x}=32$)& \cellcolor{red!25}{No difference} & 42:58 & Linear Regression \\ \hline
Iuga et. al. (2016) \cite{iuga2016baiting} & 382 & Website & & $\leq$60 &  \cellcolor{green!25}{\begin{tabular}[l]{@{}l@{}}Men performed better \end{tabular}} & 62.8:37.2& ANOVA\\ \hline
Wang et. al. (2016) \cite{wang2016overconfidence} &600 & Email &\cellcolor{red!25}{No difference} & 19-89 ($\bar{x}=52$) & \cellcolor{green!25}{\begin{tabular}[l]{@{}l@{}}Men performed better \end{tabular}}  & 36:64 & OLS Regression\\ \hline
Alsharnouby et. al. (2015) \cite{alsharnouby2015phishing} & 21 & Website &\cellcolor{red!25}{No difference} & 18-51 ($\bar{x}=27$, $\sigma=10.1$)&\cellcolor{red!25}{No difference}& 43:57& \begin{tabular}[c]{@{}c@{}}Age: Pearson Correlation\\Gender: t-test \end{tabular}\\ \hline
Wright et. al. (2014) \cite{wright2014research} & 2624 & Email & & N/A & \cellcolor{green!25}{Men performed better}&  46:54& Logistic Regression  \\ \hline
Purkait et. al. (2014) \cite{purkait2014empirical} & 621 & Website & \cellcolor{green!55}{\begin{tabular}[l]{@{}l@{}}Elders are more susceptible\end{tabular}} & 20-62 ($\bar{x}=33$, $\sigma=10$) & \cellcolor{red!25}{No difference} & 66:34& Linear Regression\\ \hline
Flores et. al. (2014) \cite{flores2014using} & 92 & Email &\cellcolor{red!25}{No difference}& N/A & \cellcolor{green!25}{Men performed better} & N/A & Point-Biserial Correlation\\ \hline
Halevi et. al. (2013) \cite{halevi2013phishing} & 100 & Email & & 18-31 ($\bar{x}=21.1$) & \cellcolor{green!25}{Men performed better} & 83:17 & Pearson's Correlation \\ \hline
Parsons et. al. (2013) \cite{parsons2013phishing} & 117 & Email & \cellcolor{red!25}{No difference} & $\leq$ 30 & \cellcolor{red!25}{No Difference} & 23:67 & \begin{tabular}[c]{@{}c@{}}Mann-Whitney U\end{tabular} \\ \hline
Wang et. al. (2012)\cite{wang2012research} & 267 & Email & \cellcolor{green!25}{\begin{tabular}[l]{@{}l@{}}Elders are less susceptible\end{tabular}} & $\bar{x}=21$, $\sigma=3.2$ & \cellcolor{red!25}{No Difference}& 46:54&PLS Regression \\ \hline
Blythe et. al. (2011) \cite{blythe2011f} & 224 & Email & \cellcolor{red!25}{No difference} & 18-65 & \cellcolor{green!25}{Men performed better} & 52:48 &  Three-way ANOVA \\ \hline
Kumaraguru et. al. (2010) \cite{kumaraguru2010teaching} & 28 & Website & \cellcolor{green!25}{13-17 years old are most susceptible} & 13-65& \cellcolor{green!25}{Men performed better} & 78:15.6& \begin{tabular}[c]{@{}c@{}}Age: ANOVA\\Gender: t-test\end{tabular}\\ \hline
Sheng et. al. (2010) \cite{sheng2010falls} & 1,001 & Email & \cellcolor{green!25}{18-25 years old are more susceptible} & $\geq$18 & \cellcolor{green!25}{Men performed better} & 48:52 &ANOVA; t-test \\ \hline
Kumaraguru et. al. (2009) \cite{kumaraguru2009school} & 515 & Email & \cellcolor{green!25}{18-25 years old are more susceptible} &\begin{tabular}[c]{@{}c@{}}18-77\\($\bar{x}=32.3$, $\sigma=12.8$)\end{tabular} & \cellcolor{red!25}{No difference} & 55.2:44.8 & \begin{tabular}[c]{@{}c@{}}Age: Pearson's $\chi^2$\\Gender: Linear Regression\end{tabular}\\\hline
Kumaraguru et. al. (2008) \cite{kumaraguru2008lessons} &311 & Email  &  & N/A & \cellcolor{red!25}{No difference} & 58:42 & t-test\\ \hline
Herzberg and Jbara (2008) \cite{herzberg2008security} &23 & Website & \cellcolor{red!25}{No difference} & 19-50 ($\bar{x}=25$) & \cellcolor{red!25}{No difference} &  82.7:17.3 & N/A \\\hline
Martin (2008) \cite{martin2008} & 141 & Email & \cellcolor{green!25}{Elders are less susceptible} & $\geq18$ & \cellcolor{red!25}{No difference} & 60:40 & Pearson Correlation\\\hline
Jagatic et. al. (2007) \cite{Jagatic2007} & 37,821 & Email & & 18-24& \cellcolor{green!25}{Men performed better} & 48:52 & t-test \\\hline
Sheng et. al. (2007) \cite{sheng2007anti}& 42 & Website & \cellcolor{red!25}{No difference} & 18-34 & \cellcolor{red!25}{No difference}& 14:86 &Spearman Correlation \\\hline
Kumaraguru et. al. (2007b) \cite{kumaraguru2007getting} & 42 & Email & \cellcolor{red!25}{No difference}&  $\bar{x}=25.6$, $\sigma=8.13$& \cellcolor{red!25}{No difference} & 17:83 & \begin{tabular}[c]{@{}c@{}}Age: Pearson's $\chi^2$\\Gender: t-test\end{tabular}\\ \hline
Dhamija et. al. (2006) \cite{dhamija2006phishing} &22 &Website& \cellcolor{red!25}{No difference} & 18-56 ($\bar{x}=29.9$, $\sigma=10.8$) & \cellcolor{red!25}{No difference} & 45:55 & \begin{tabular}[c]{@{}c@{}}Age: Pearson's $\chi^2$\\Gender: t-test\end{tabular} \\\hline
\end{tabular}
}
\label{table:age_gender}
\end{sidewaystable*}

\subsection{Individual Attributes}
\hl{A total of 44 papers covered all or a subset of individual attributes. Since there was a high overlap between papers that studied age and gender, we report their results together and then cover the studies on background knowledge.}

\subsubsection{Age and Gender}
\label{sec:sub_gender_effect}

Table \ref{table:age_gender} summarizes the findings of 35 previous studies on the difference between the attack detection ability of Internet users based on their age and gender. The papers are sorted in reverse chronological order from top to bottom. An empty cell means that the variable has not been studied. The \textit{Vector} column shows the attack vector (email or website) given to the participants for the experiments conducted. 
The last column shows the statistical significance tests used. The reason for combining the results of studies that analyzed both age and gender in a single table is the high overlap  between the studies that analyzed both age and gender. There was one study that studied gender and age combined together \cite{lin2019susceptibility}. The authors recruited two different groups of participants to analyze the age effect (old and young) so we reported gender and age distributions of each sample set separately in the table. 

For \textit{age}, unlike the common belief, 14 out of 26 works that studied age found no statistically significant relationship between participants' vulnerability and their age. Nine of the reported studies found younger people more vulnerable while two found the opposite (one study mixed the age and gender together and found older women the most susceptible). However, the definitions of ``elder'' and ``younger'' change based on the statistical tests used. Most studies used correlation tests or regression analysis to evaluate the relation between age and phishing vulnerability with five exceptions. 
Authors in \cite{blythe2011f,kumaraguru2010teaching,sheng2010falls,valipe2018} divided the participants based on their age into three, four, five, and six ranges respectively and used Analysis of Variance (ANOVA) to compare the performance between the age groups. Researchers in \cite{parsons2013phishing} did not report the number of ranges or thresholds that they used.


The results for \textit{gender} are  similar to those for  \textit{age}. Most of the studies (24 out of 35) found no significant difference between males and females in their ability to detect phishing and legitimate emails or websites. Nine studies found females more susceptible and one study found the opposite (males are more susceptible).

\begin{figure}
    \centering
    \captionsetup{justification=centering}
    \includegraphics[width=\columnwidth]{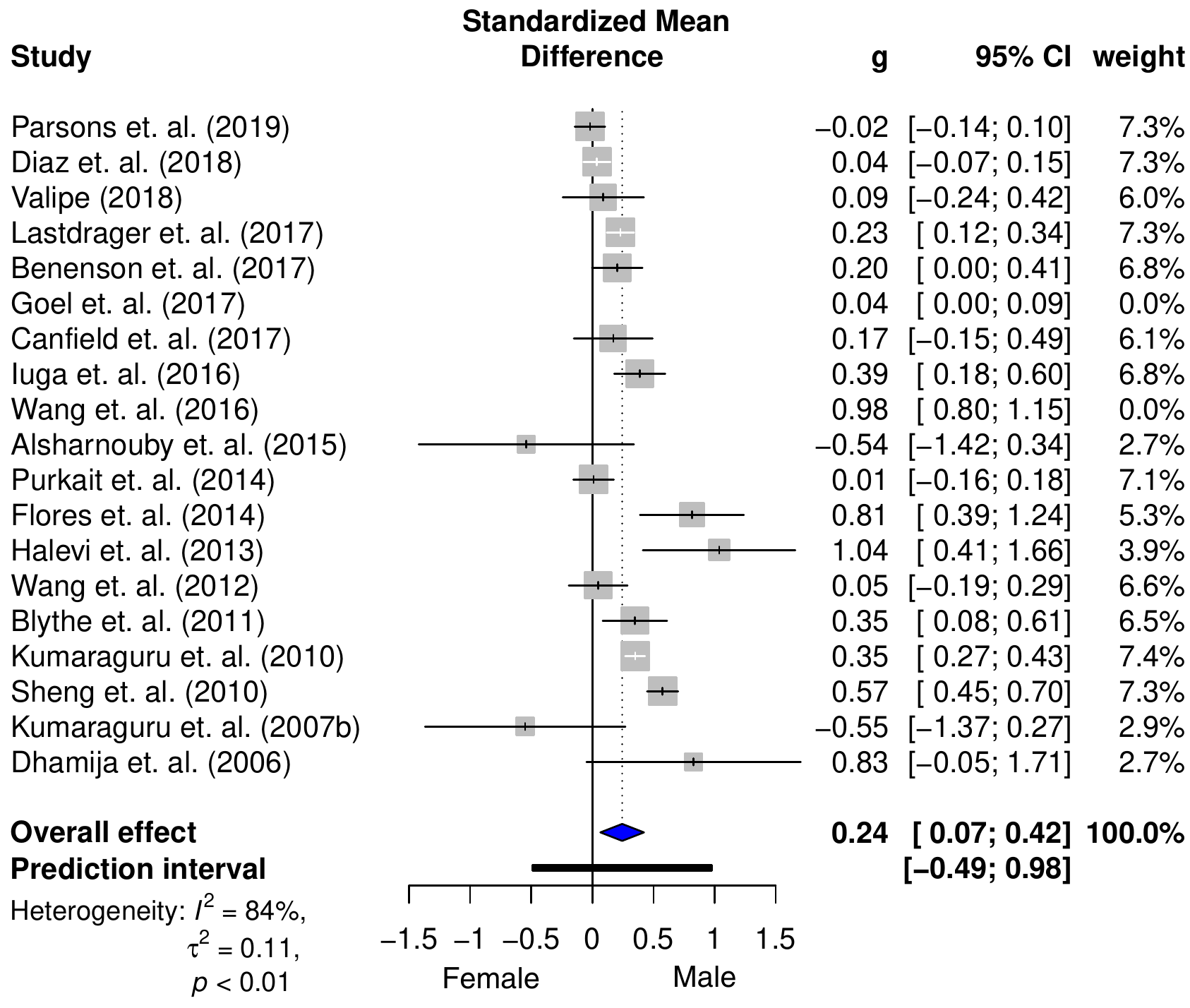}
    \caption{Forest plot of the effect size of \textit{gender} on phishing detection ability}~\label{figure:gender_forest}
\end{figure}

\begin{figure}
    \centering
    \includegraphics[width=\columnwidth]{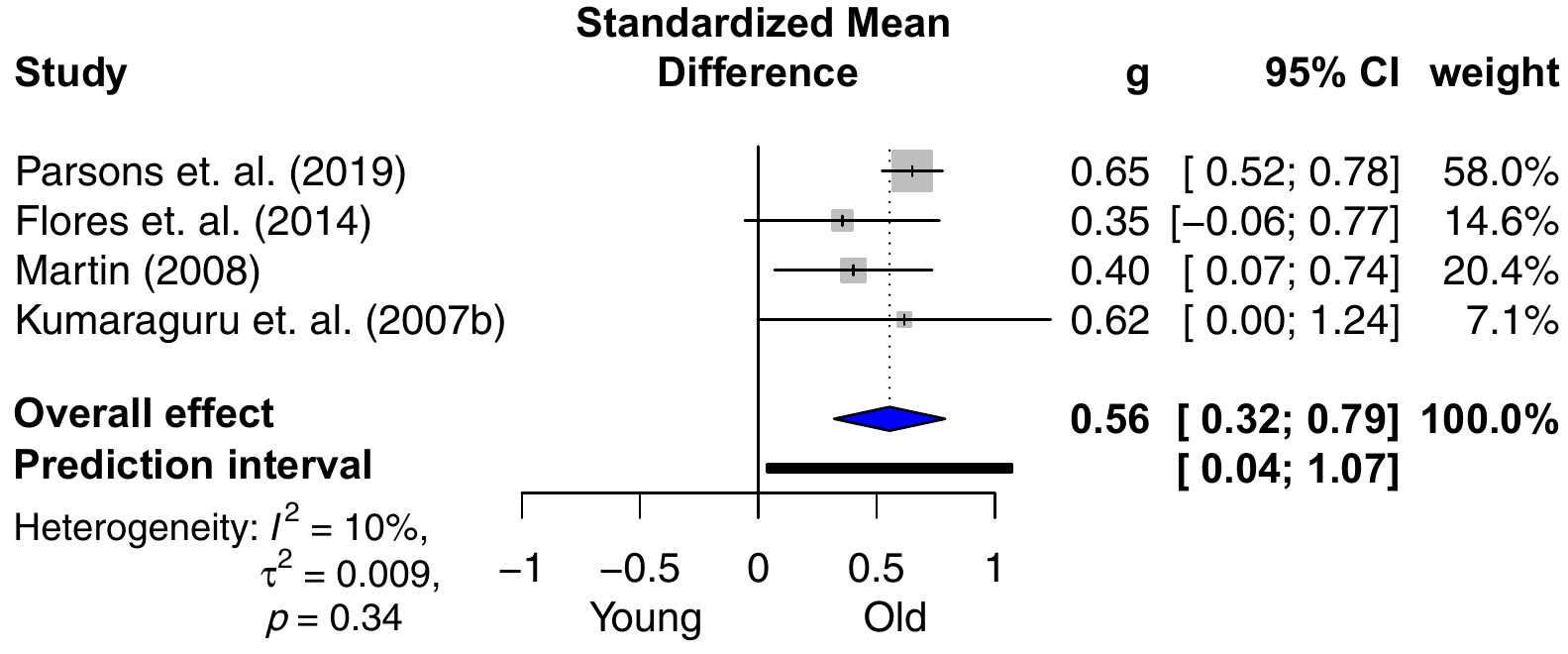}
    \caption{Forest plot of the effect size of \textit{age} on phishing detection ability}~\label{figure:age_forest}
\end{figure}

Due to the contradictory results observed from previous studies regarding the effect of gender and age, meta-analysis can be helpful to assess their results. It can help us to provide a more precise estimate of the age and gender effect \cite{haidich2010meta}. After analyzing the reported results of the studies on age and gender, we realized that several failed to report their findings in a way so that we can calculate the effect size. The most frequent issues that we encountered were: 1) not reporting the sizes of the control and intervention groups, 2) only reporting the p-values in t-tests instead of reporting the $t$ value, 3) not reporting the mean and standard deviation of the dependent variable when using regression analysis, and  4) not mentioning the codings used for different groups (e.g. when comparing males and females in regression, it is not clear which value (0 or 1) has been used for each category). We were able to calculate the effect sizes for 19 studies on  gender and only six studies on age.

Figures \ref{figure:gender_forest} and \ref{figure:age_forest} show the forest plots of the Random Effects Models using Hedge's $g$ \cite{hedges1984nonparametric} to measure the effect size for gender and age respectively. We used the Random Effects Model instead of a Fixed Effects Model since it performs better when studies use different populations \cite{schwarzer2015meta}. After applying outlier and influence diagnostics \cite{viechtbauer2010outlier}, four studies (\cite{wang2016overconfidence, goel2017got, diaz2020phishing, sheng2007anti}) have been removed from the models. The overall effect shows that the difference between the performance of male and female is significantly different (males performed better, $t = 2.9, p < 0.05$). For the \textit{age} also, the meta-analysis results shows a significant relationship (Older users performed better, $t=7.53, p<0.01$). As mentioned earlier, researchers used different methods to study the effect of \textit{age}, some used correlation/regression and some grouped participants based on their age and compared the performance between the groups. Combining these studies might affect the validity of the meta-analysis. Fortunately, all the studies included in the meta-analysis of \textit{age} used correlation or regression (except \cite{diaz2020phishing} which was detected as an outlier). 

The random effects model for \textit{gender} has a significant level of heterogeneity ($I^2=84\%, p<0.01$ ), which can occur because of several reasons, e.g. population, vector of study, and study design. 

    \noindent\textbf{Population}: Typically, studies use one of these four populations:  1) students (university or high school), 2) employees of a company, 3) Amazon Mechanical Turk, and 4) University staff and students. The differences between these populations can  bias the results of the studies. We calculated the pooled effect for each subgroup but we did not find a significant difference between subgroups ($Q=2.30, df=3, p=0.512$).
    
    \noindent\textbf{Location}: \hl{Nationality of the participants is also another variable that could increase the heterogeneity of the study. There were three main types in the 37 studies that we analyzed: 1) U.S. (21 studies), 2) Five studies use public pool of participants, i.e., people from all the world could participate (e.g., Mechanical Turk), 3) Seven studies recruited their participants from countries other than U.S. (e.g., India, U.K., Germany, Australia, etc.). The reason that we grouped countries other than U.S. together was to have enough samples in that group to calculate the pooled effect size. Four studies did not mention where the participants are recruited from, so we removed them from subgroup analysis. Studies that used open public pool showed larger effect size than other studies (Hedges' $g$: public=$0.41$, US=$0.24$, and Other=$0.08$), but their difference was only marginally significant ($Q=5.75, df=2, p=0.056$) }
    
    \noindent\textbf{Vector}: Some studies used email instances to evaluate users susceptibility and some used websites. This can affect their findings since they calculate different outcomes variables. We created the Random Effects models using the studies on each vector but the heterogeneity was still high. We did not find any significant difference between studies using emails, websites, or both ($Q=0.09, df=2, p=0.955$).
    
    \noindent\textbf{Study Design}: Even if two studies use only emails (website) for testing susceptibility, there may be a huge difference between them, since they usually create their own set of emails (websites) and this can result in some unaccounted variables affecting the findings. \hl{Authors in} \cite{lin2019susceptibility} \hl{showed that different persuasive techniques used by attackers (reciprocation, liking, scarcity, etc.) result in different level of susceptibility. Based on the emails/websites collected by different researchers, only a subset of these techniques will exist in each study and that could create a big difference in the outcome of the study.} Besides this, the method of delivering the attacks to participants (real-world attack versus role-playing) is another factor in the heterogeneity of the studies.

Publication bias is another common issue in research that may seriously alter the estimate of the effect under investigation. Publication bias states that studies with high effect sizes are more likely to be published than studies with low effect sizes \cite{rothstein_sutton_borenstein_2005}. A common way to investigate potential publication bias in a meta-analysis is the funnel plot. In the funnel plot, the X-axis shows the effect size and the Y-axis represents the standard error. So, the larger studies (smaller standard error) would appear next to the Y-axis and closer to zero. The dotted funnel is centered around the pooled effect size. In the case of no publication bias, all studies appears symmetrically around the pooled effect size. However, if publication bias exists, the funnel should look asymmetric since among the small studies (higher standard error) only those with large effect size are published (small studies with no significant effect size would be missing). The colored funnel is centered around the zero effect size (i.e., at the value under the null hypothesis of no effect) and highlights the regions with different significant levels (white area shows the studies that are not significant). 

\begin{figure}
    \centering
    \includegraphics[width=0.95\columnwidth]{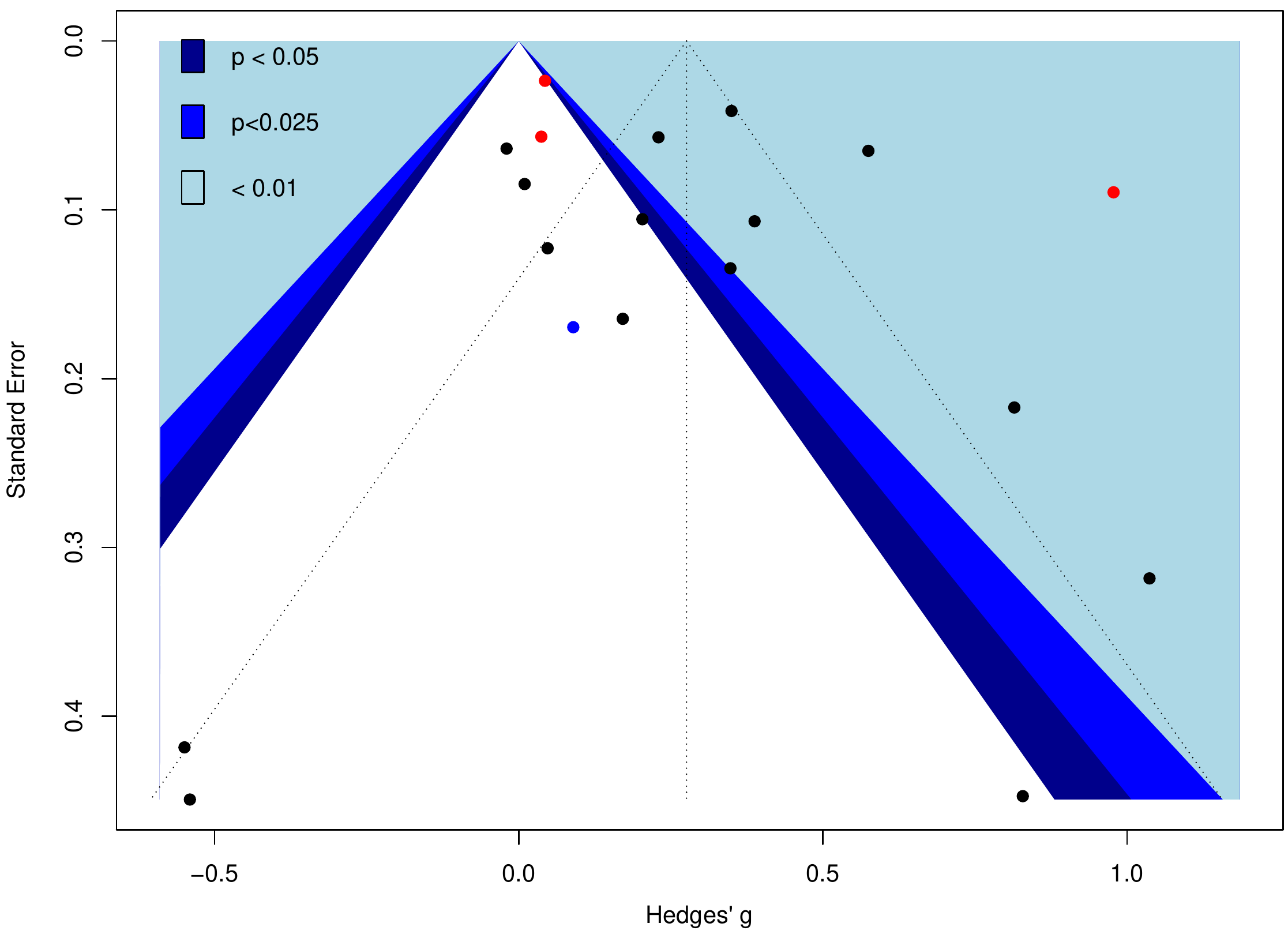}
    \caption{Funnel plot of the effect of gender on phishing detection ability (female versus male). Red points represent outliers and blue points unpublished works (dissertation)}~\label{figure:gender_funnel}
\end{figure}

Of the variables studied here, we only report the funnel plot for variables with at least 10 studies (with fewer studies the power of the tests is too low to distinguish chance from real bias \cite{higgins_thomas_2019}). As mentioned earlier, the asymmetrical distribution of studies in the funnel plot indicates potential publication bias.
Figure \ref{figure:gender_funnel} shows the funnel plot of the 19 studies, which were included in the meta-analysis, on \textit{gender} effect. Red points show the studies which are detected as outliers and the blue point shows the dissertation (unpublished work). We used Egger’s test of the intercept \cite{egger1997bias} to quantify the funnel plot asymmetry. The p-value of Egger’s test shows no significant asymmetry (p=0.18). So, based on the funnel plot's result, there is no publication bias among studies that we analyzed.
 
\subsubsection{Background Knowledge}
Table \ref{table:background} presents the observations for background knowledge (sorted in reverse chronological order). Since there are many different ways of measuring the background knowledge of participants, we also report how it was computed. Unlike Table \ref{table:age_gender}, the findings are more consistent here. Most of the studies that asked participants to differentiate between phishing and legitimate emails/websites found background knowledge as a positive factor. Six studies that did not find it significant.
Surprisingly, authors in \cite{parsons2013phishing} found that participants who had formal training on information system or information technology in the past perform worse than those who did not have the training. This finding is the exact opposite of what researchers found in \cite{diaz2020phishing,sheng2010falls}. 
Unfortunately, authors in \cite{parsons2013phishing} did not report the number of participants who had the formal training, so the opposite finding could be due to lack of enough samples in one group. Other reasons could be the period of the training (training was too far in the past) or quality of the training (in terms of content and length).

\begin{table*}
\caption{Studies that compared users based on their background/technological knowledge (sorted based on their publication year). Green cells show papers with significant findings (brighter green shows opposite findings). Gender distribution column only shows male and female percentage (male:female). SEM: Structural Equation Modeling.}
\resizebox{\textwidth}{!}{
\begin{tabular}{|c|r|c|c|c|c|}
\hline
\textbf{Paper} & \textbf{Size} & \textbf{Vector} & \textbf{Background Knowledge Measure} & \textbf{Gender dist.} & \textbf{Statistics} \\ \hline

\cellcolor{red!25}{\begin{tabular}[l]{@{}c@{}}Williams and\\Polage '19 \cite{williams2019persuasive}\end{tabular}} & 178 & Email & Email habitual use has no effect& 24:76 & Bivariate Correlation \\ \hline
\cellcolor{green!25}{Parsons et. al. '19 \cite{parsons2019predicting}} & 985 & Email & Time spent on computer improves performance & 53:47 & Hierarchical Regression \\ \hline
\cellcolor{green!25}{Musuva et. al. '19 \cite{musuva2019new}} & 192 & Email & Knowledge of threat improves performance & 63:37 &  SEM\\ \hline
\cellcolor{green!25}{Taib et. al. '19\cite{taib2019social}} & 56,365 & Email & Employment history (New employees are worst) & 49.6:50.4& Pearson's $\chi^2$\\ \hline
\cellcolor{green!25}{Orunsolu et. al. '18 \cite{orunsolu2018users}} & 427 & Both & Higher computer/Internet knowledge improves performance  & 57:43 & Linear Regression\\ \hline
\cellcolor{green!25}{Diaz et. al.'18 \cite{diaz2020phishing}} & 1,350 & Email &\begin{tabular}[c]{@{}c@{}} Users with previous training and \\senior students performed better\end{tabular} & N/A & Pearson's $\chi^2$ \\ \hline
\cellcolor{red!25}{Valipe '18 \cite{valipe2018}} & 156& Email &  Employment history has no effect & 59:36 & ANOVA \\\hline
\cellcolor{green!25}{Lastdrager et. al. '17 \cite{lastdrager2017effective}}& 353 & Both & Having email/FB account improves performance & 46:54 & t-test\\\hline
\cellcolor{green!25}{Jensen et. al. '17\cite{jensen2017combating}} & 104 & Email & Having higher computer self efficacy improves performance & 72:28 & ANCOVA \\ \hline
\cellcolor{green!25}{Jensen et. al. '17\cite{jensen2017training}} &355 & Email & Faculty members perform better than students & N/A &Logistic Regression\\ \hline
\cellcolor{green!25}{Bullee et. al. '17 \cite{bullee2017spear}} & 593 & Email & Having longer employment history improves performance & 75.5:24.5 & Logistic Regression\\ \hline
\cellcolor{green!25}{Butavicius et. al. '17\cite{butavicius2017understanding}} & 121 & Email & Having higher security awareness improves performance & 32:68& Linear Regression \\ \hline
\cellcolor{green!25}{Moody et. al. '17 \cite{moody2017phish}} & 595 & Email & Users with higher Internet usage perform better & 56:44& Logistic Regression\\ \hline
\cellcolor{green!25}{Iuga et. al. '16 \cite{iuga2016baiting}} & 382 & Email & Users with higher PC usage perform better & 62.8:37.2 & ANOVA\\ \hline
\cellcolor{green!25}{Li et. al. '16 \cite{li2016examination}} & 592 & Email & Users who have more familiarity with email perform better & N/A & Linear Regression \\ \hline
\cellcolor{green!25}{Harrison et. al. '16 \cite{harrison2016}} & 194 & Email & Users with more email experience perform better& 52.8:47.2 & Logistic Regression\\ \hline
\cellcolor{green!25}{Vishwanath '15 \cite{vishwanath2015examining}} & 192 & Email & Users with habitual email use perform better & N/A & Logistic Regression\\ \hline
\cellcolor{red!25}{Alsharnouby et. al. '15 \cite{alsharnouby2015phishing}} & 21 & Website & Technical proficiency score has no effect & 43:57 & Pearson's Correlation\\ \hline
\cellcolor{green!55}{Parsons et. al. '13\cite{parsons2013phishing}} & 117 & Email & Users who had previous training perform worse & 23:77 & t-test \\ \hline
\cellcolor{green!25}{Pattinson et. al. '12\cite{pattinson2012some}} & 117 & Email & Users with higher computer familiarity perform better & N/A&  Spearman's Correlation\\ \hline
\cellcolor{green!25}{Wright and Marret '10 \cite{wright2010influence}} & 299 & Email & User with higher web experience (self-reported) perform better& 63:37 &  SEM\\ \hline
\cellcolor{green!25}{Sheng et. al. '10\cite{sheng2010falls}} & 1,001 & Email & Users who had previous training perform better & 48:52 & t-test\\ \hline
\cellcolor{red!25}{Sheng et. al. '07\cite{sheng2007anti}} & 42& Website & Time spent on computer has no effect  & 14:86 & Spearman's Correlation\\\hline
\cellcolor{green!25}{Downs et. al. '07\cite{downs2007behavioral}} & 232 & Email & Having higher knowledge of phishing improves performance &  N/A & Pearson's $\chi^2$\\\hline
\cellcolor{red!25}{Kumaraguru et. al. '07b\cite{kumaraguru2007getting}} & 42 & Email  & Internet usage has no effect& 17:83 & Pearson's Correlation\\\hline
\cellcolor{red!25}{Dhamija et. al. '06\cite{dhamija2006phishing}} & 22 & Website & Time spent on computer has no effect & 45:55 & Pearson's Correlation\\\hline
\end{tabular}
}
\label{table:background}
\end{table*}

Authors in \cite{sheng2010falls}, using mediation analysis, showed that women have less technical experience than men and therefore they are more susceptible  to phishing. So, we added the ``gender distribution'' column to Table \ref{table:background} to see if there is any bias toward a specific gender or not. 
To investigate this finding, we can compare the papers in Tables \ref{table:age_gender} and \ref{table:background} that studied the effect of both gender and background knowledge on phishing susceptibility. We can categorize these works into two groups: 1) studies that showed participants with higher background knowledge perform better while there is no difference in performance between male and female \cite{parsons2019predicting, musuva2019new} 2) studies that showed there is no relationship between background knowledge as well as gender and phishing susceptibility \cite{valipe2018}. The findings of both groups do not support conclusion of \cite{sheng2010falls} that women are more susceptible due to lower technical experience.

\begin{figure}
    \centering
    \includegraphics[width=\columnwidth]{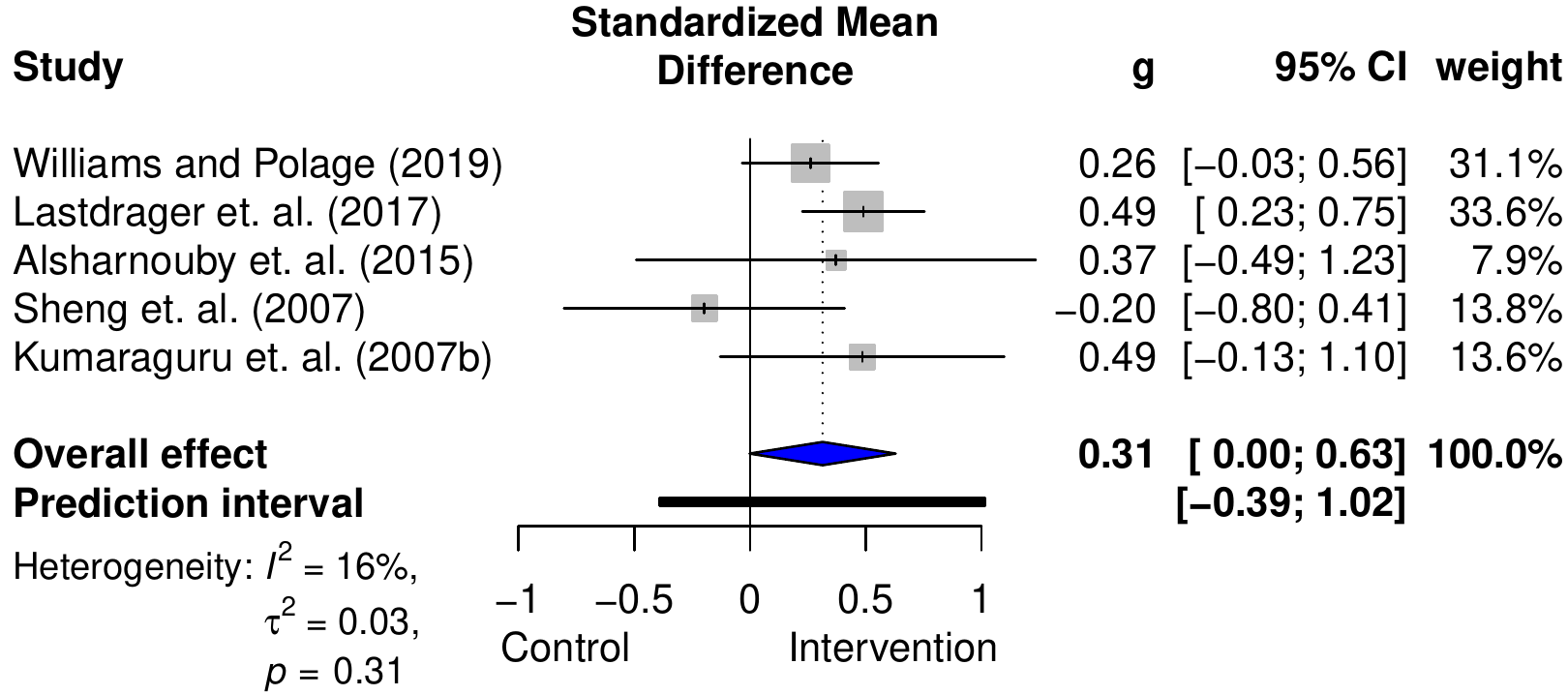}
    \caption{Forest plot of the effect size of \textit{background knowledge} on phishing detection ability}~\label{figure:background_forest}
\end{figure}

We could only measure the effect sizes for six studies (Figure \ref{figure:background_forest}) and the Random Effect Model shows that increase in background knowledge decreases phishing susceptibility, but it is marginally  significant ($t=2.76, p=0.051$). One study \cite{diaz2020phishing} has been removed from the model since it was detected as highly influential on heterogeneity (does not have a large impact on the overall results, but it does add substantially to the heterogeneity found in the meta-analysis \cite{viechtbauer2010outlier}).



\subsection{External attributes}
\hl{As mentioned earlier, fewer papers covered external attributes compared to the individual attributes. There were 42 papers that studied external attributes.}

\subsubsection{Training}
Training users on how to detect phishing attacks is expected to improve their detection ability. Almost all of the studies support this hypothesis (Table \ref{table:training}) with some observations regarding the type of users and the recency of training.
Researchers in \cite{cj2018phishy, kumaraguru2010teaching} divided the participants based on their performance (before delivering the training) into three categories (novice, intermediate, and expert). Their results showed that the training significantly improved the performance of the novice and intermediate participants but did not help experts. 
Studying the effect of training over the course of time has been done in \cite{lastdrager2017effective, nyeste2011,  dodge2012empirical, kumaraguru2010teaching} but their results contradict each other. 
Authors in \cite{lastdrager2017effective} tested participants 0/2/4 weeks after training and showed that although users' performance improved after the training, its effect disappeared after four weeks. Also, authors in \cite{nyeste2011} showed that people are more likely to be phished in the second week than the first week. However, researchers in \cite{dodge2012empirical} observed the opposite, the training was not effective after 10 days of delivery, but it was effective after 63 days.
No difference was observed in \cite{kumaraguru2010teaching} as they tested immediately and one week after training. 
The difference in the population of these studies could be a factor in these different results. Authors in \cite{lastdrager2017effective} used pupils from primary and elementary schools while a general population was used in \cite{kumaraguru2010teaching} and university students in \cite{nyeste2011}. More research is needed to develop a better understanding of the effect of training over a period of time.

\begin{sidewaystable*}
\caption{Studies that analyzed the effect of training on phishing susceptibility (sorted based on their publication year). Green cells show significant findings. Orange cells show partial improvement, which is explained in the Training Type column.  Gender distribution column only shows male and female percentage (male:female), so if an study has a third category (e.g. others) their sum does not add up to 100 (\cite{cj2018phishy, kumaraguru2010teaching}).}
\begin{threeparttable}
\resizebox{\textwidth}{!}{
\begin{tabular}{|c|r|c|c|c|c|c|c|}
\hline
\textbf{Paper} & \textbf{Size} & \textbf{Vector} & \textbf{Training Type} & \textbf{Amount} & \textbf{Gender dist.} & \textbf{Age Range} & \textbf{Statistics} \\ \hline
\cellcolor{green!25}{Martin '19\cite{martin2019}} & 115 & Email & Showed the recorded video of the presentation & N/A & 62.3:37.7 &  N/A & t-test\\\hline
\cellcolor{green!25}{Wen et. al. '19 \cite{wen2019hack}} & 39 & Email & Game (improved performance) & N/A & N/A & N/A & ANOVA\\ \hline
\cellcolor{green!25}{Harrison et. al. '19\cite{harrison2019learning}} & 442 & Email & \begin{tabular}[l]{@{}c@{}}An instructor taught the material (instructor-based)\\(improved performance) \end{tabular}& 15 mins & N/A & N/A & ANOVA\\\hline
\cellcolor{orange!35}{\begin{tabular}[l]{@{}c@{}}Wash and\\Cooper '18\cite{wash2018provides}\end{tabular}} & 1945 & Email & \begin{tabular}[c]{@{}c@{}}Embedded training;\tnote{a} ~ Performance Improved when the narrative\\ comes from a security expert (not a peer)
\end{tabular}& Four emails & N/A & N/A & $\chi^2$ test\\ \hline
\cellcolor{orange!35}{CJ et. al. '18\cite{cj2018phishy}} & 8071 & Website & Game (only novice and intermediate users improved) & N/A & 62.2:35.1 & 21-50& t-test\\ \hline
\cellcolor{green!25}{Gordon et. al. '18\cite{gordon2018employee}} & N/A & Email & Repeated phishing campaigns reduced click rate & N/A& N/A & N/A & Logistic Regression \\ \hline
\cellcolor{orange!35}{McElwee et. al. '18 \cite{mcelwee2018influencing}} & N/A & Email & \begin{tabular}[c]{@{}c@{}}Simulated phishing campaign; \\Targeted training improved but not generic training \end{tabular}& N/A & N/A & N/A & \begin{tabular}[c]{@{}c@{}}Linear Regression\\t-test \end{tabular}\\\hline
\cellcolor{green!25}{Lastdrager et. al. '17 \cite{lastdrager2017effective}} & 353 & Both &\begin{tabular}[c]{@{}c@{}} Interactive presentation where participants could ask\\ question during the presentation (improved performance) \end{tabular}& 40 mins & 46:54 & \begin{tabular}[c]{@{}c@{}}8-13 \\($\bar{x}=10.6, \sigma=1.0$) \end{tabular} & t-test\\ \hline
\cellcolor{green!25}{Jensen et. al. '17\cite{jensen2017training}} & 355 & Email & Comic and Text (both improved performance) & N/A & N/A & $\geq$18 & Logistic Regression \\ \hline
\cellcolor{green!25}{\begin{tabular}[c]{@{}c@{}}Moreno-Fern{\'a}ndez\\et. al. '17 \cite{moreno2017fishing}\end{tabular}} & 175 & website &\begin{tabular}[c]{@{}c@{}}Showing phishing and their target websites side by side;\\Easy-to-hard training performed better than hard-to-easy\tnote{b}\end{tabular}&N/A &  65:35 & \begin{tabular}[c]{@{}c@{}}18-66 \\($\bar{x}=38$, $\sigma=9.6$)\end{tabular} & \begin{tabular}[c]{@{}c@{}}Mann-Whitney U\end{tabular} \\ \hline
\cellcolor{green!25}{Xiong et. al. '17\cite{xiong2017domain}} & 320 & Website & Training to focus on links improved performance & N/A & 42:58 & $\geq$18 & $\chi^2$ test\\ \hline
\cellcolor{orange!35}{Yang et. al. '17 \cite{yang2017use}} & 63 & Email & \begin{tabular}[l]{@{}c@{}}Instructor-based training alone is not helpful;\\Training combined with warning improved the performance\end{tabular} & 10 mins & 54:46 & 19-30 & $\chi^2$ test \\ \hline
\cellcolor{green!25}{Stockhardt et. al. '16\cite{stockhardt2016teaching}} & 81 & Website & \begin{tabular}[l]{@{}c@{}}Instructor-based; computer-based; text-based;\\Computer-based training was the best \end{tabular}& Instruc.: 45 mins&  83:17 & $\bar{x}=19.6$, $\sigma=4.4$ & ANOVA\\ \hline
\cellcolor{green!25}{Arachchilage et. al. '16\cite{arachchilage2016phishing}}& 20 & website & Game (improved performance) & 10 mins & 65:35 & 18-15 & t-test \\ \hline
\cellcolor{green!25}{Kunz et. al. '16\cite{kunz2016nophish}} & 32 & Website &  Online Material (improved performance) & N/A & 56:44 & \begin{tabular}[c]{@{}c@{}}19-56\\ ($\bar{x}=28.5$, $\sigma=10.8$) \end{tabular}& t-test\\ \hline
\cellcolor{green!25}{Arachchilage '15\cite{arachchilage2015can}} & 40 & Website & \begin{tabular}[l]{@{}c@{}}Game; APWG Training;\\ Game made improvements but APWG did not\end{tabular} & 10 mins & 65:35 & 18-25 &t-test\\ \hline
\cellcolor{red!25}{Jones '15 \cite{jones2015}} & 91 & Email & Game (no improvement) & 15 mins & 27:73 & 18-64  & ANOVA\\\hline
\cellcolor{green!25}{Scott et. al. '14 \cite{scott2014assessing}} & 66 & Website & Game (improved performance) & N/A & N/A & $\geq$18 (undergrads) & ANCOVA  \\ \hline
 \cellcolor{red!25}{House '13\cite{house2013}} & 101 & Email & Training Video (no improvement) & $\bar{x}=166$ sec& 50.5:49.5 & 18-59 & Logistic Regression\\\hline
\cellcolor{green!25}{Yang et. al. '12\cite{yang2012building}} & 62 & Website & Game (improved performance)& N/A & N/A & N/A & t-test  \\\hline
\cellcolor{orange!35}{Dodge et. al. '12\cite{dodge2012empirical}} & 892 & Email &  \begin{tabular}[l]{@{}c@{}} Institution's phishing awareness training\\Short period (10 days): not effective\\Long period (63 days): improved\end{tabular} &N/A & N/A & 18-26 & ANOVA\\ \hline
\cellcolor{green!25}{Nyeste '11 \cite{nyeste2011}} & 84 & Email & \begin{tabular}[l]{@{}c@{}}Comic; Game\\ Both improved performance  \end{tabular}& N/A & 65:35 & N/A & ANOVA\\ \hline
\cellcolor{orange!35}{Kumaraguru et. al. '10\cite{kumaraguru2010teaching}} & 28& Website &\begin{tabular}[l]{@{}c@{}} Different online materials \\ Only novice and intermediate users showed improvements \end{tabular} & One to five pages& 78:15.6 & 13-65& t-test\\\hline
\cellcolor{green!25}{Sheng et. al. '10\cite{sheng2010falls}} & 1,001 & Email & \begin{tabular}[l]{@{}c@{}}Cartoon; Game; Online Materials \\ All methods improved performance\end{tabular} &\begin{tabular}[l]{@{}c@{}}Cartoon: 0.5 min\\Game: 8 mins\\Online: 1.5 mins\end{tabular} &48:52 & $\geq$18 &  t-test\\\hline
\cellcolor{green!25}{Kumaraguru et. al. '09 \cite{kumaraguru2009school}} & 515 & Email &\begin{tabular}[l]{@{}c@{}} Embedded Training\tnote{a} \\Retain knowledge after 2days/1week/2weeks\end{tabular} & N/A&55.2:44.8 & \begin{tabular}[c]{@{}c@{}}18-77\\ ($\bar{x}=32.3$, $\sigma=12.8$) \end{tabular}& ANOVA\\\hline
\cellcolor{green!25}{Alnajim and Munro '09\cite{alnajim2009anti}} & 36 & Website & \begin{tabular}[l]{@{}c@{}}Embedded Training\tnote{a} ~(improved performance)\end{tabular} & 1 email out of 14 & 33.3:66.7 & 18-39 & ANOVA\\\hline
\cellcolor{green!25}{Kumaraguru et. al. '08\cite{kumaraguru2008lessons}} & 311 & Email & \begin{tabular}[l]{@{}c@{}}Embedded Training\tnote{a} \\ Improved performance after one week\end{tabular} & 1 email out of 4 &    58:42 & N/A & t-test\\\hline
\cellcolor{orange!35}{Kumaraguru et. al. '07a \cite{kumaraguru2007protecting}} & 30 & Email & \begin{tabular}[l]{@{}c@{}}Embedded Training\tnote{a} ~(comic and text)\\ Comic based improved, text based did not\end{tabular} & 2 emails out of 19 & 37:53 & $\bar{x}=26.6$ & $\chi^2$ test\\\hline
\cellcolor{green!25}{Sheng et. al. '07\cite{sheng2007anti}} & 42 & Website & \begin{tabular}[l]{@{}c@{}}Game; Game Transcript; Online Materials \\Game made the highest improvement \end{tabular}& 15 min each & 14:86 & 18-34 & ANOVA\\\hline
\end{tabular}
}
\begin{tablenotes}
\item[a] Phishing emails are sent to users (e.g. employees of a company) from the administrators and the training will be shown if users click on the link inside the email.
\item[b] Participants are trained to detect the difference between the font of phishing and targeted legitimate websites. Differences were in some cases harder and in others\\ easier to notice. They changed the order of training materials, sometimes showing the hardest mismatches first and sometimes the  easiest.
\end{tablenotes}
\end{threeparttable}
\label{table:training}
\end{sidewaystable*}

\begin{figure}
    \centering
    \includegraphics[width=\columnwidth]{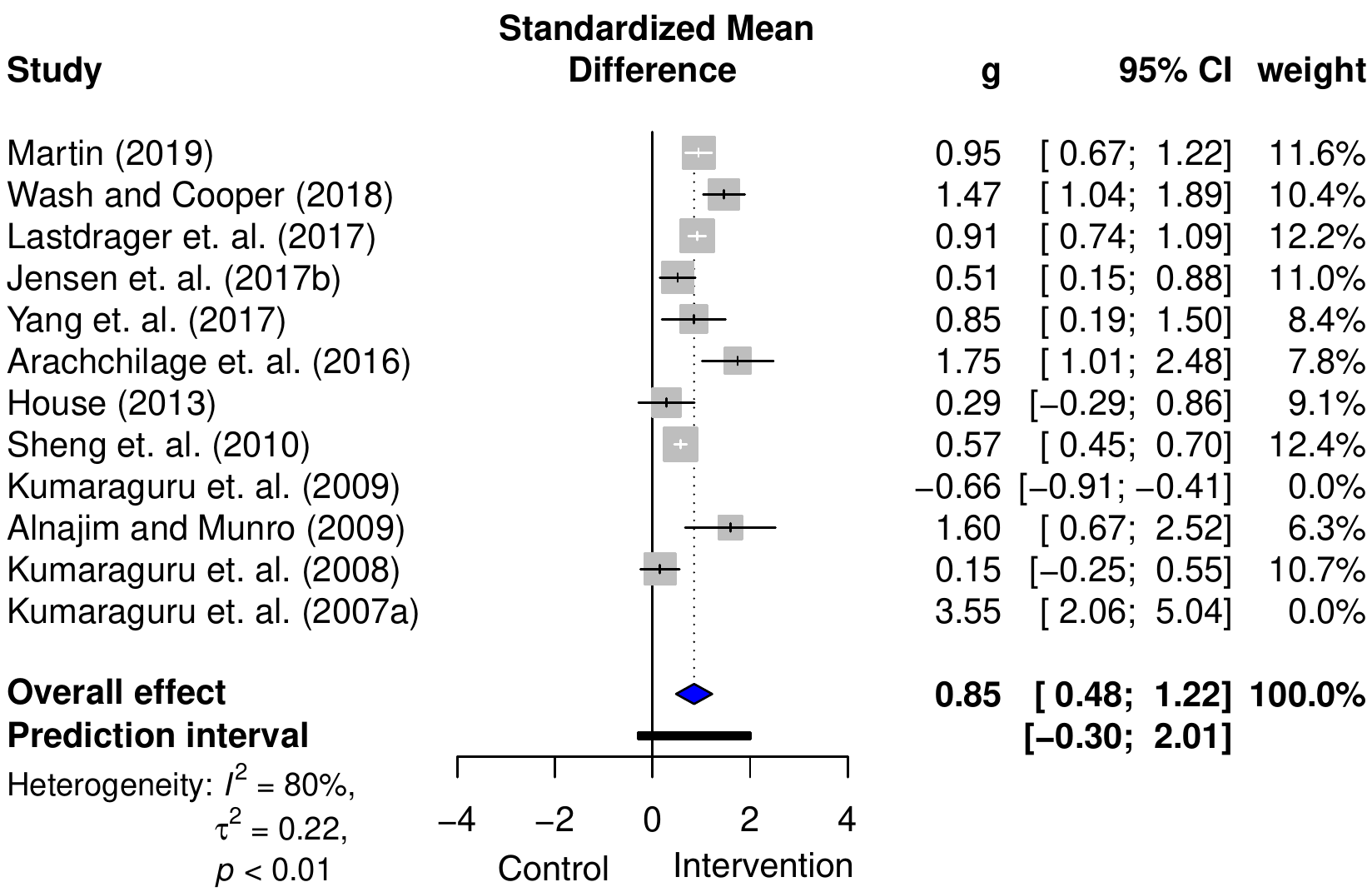}
    \caption{Forest plot of the effect size of \textit{training} on phishing detection ability}~\label{figure:training_forest}
\end{figure}

Figure \ref{figure:training_forest} shows the forest plot of the studies on the effect of training. Here also several studies failed to report their results properly, and only 12 of them are included in the meta-analysis. Lack of reporting the size of each group (control and intervention) was the main reason that prevented us from calculating the effect size. Two studies \cite{kumaraguru2009school,kumaraguru2007protecting} were detected as outliers and removed from the model (we used the method proposed in \cite{viechtbauer2010outlier} to detect outliers). Based on the findings of the remaining 10 studies, training improves detection ability significantly ($t=5.21, p<0.001$). 

Heterogeneity for the model is also high, possibly because of the same biases mentioned in Section \ref{sec:sub_gender_effect}. Subgroup analysis showed a significant difference between training effect on email and website ($Q=59.10, df=2, p<0.0001$) with training having a higher effect size on website detection (Hedges' $g$: website: 1.68, email: 0.68). \hl{However, subgroup analysis of populations and location showed no significant difference  (population: $Q=4.67, df=5, p=0.457$, location: $Q=1.88, df=2, p=0.389$)}. 

\begin{figure}
    \centering
    \includegraphics[width=0.9\columnwidth]{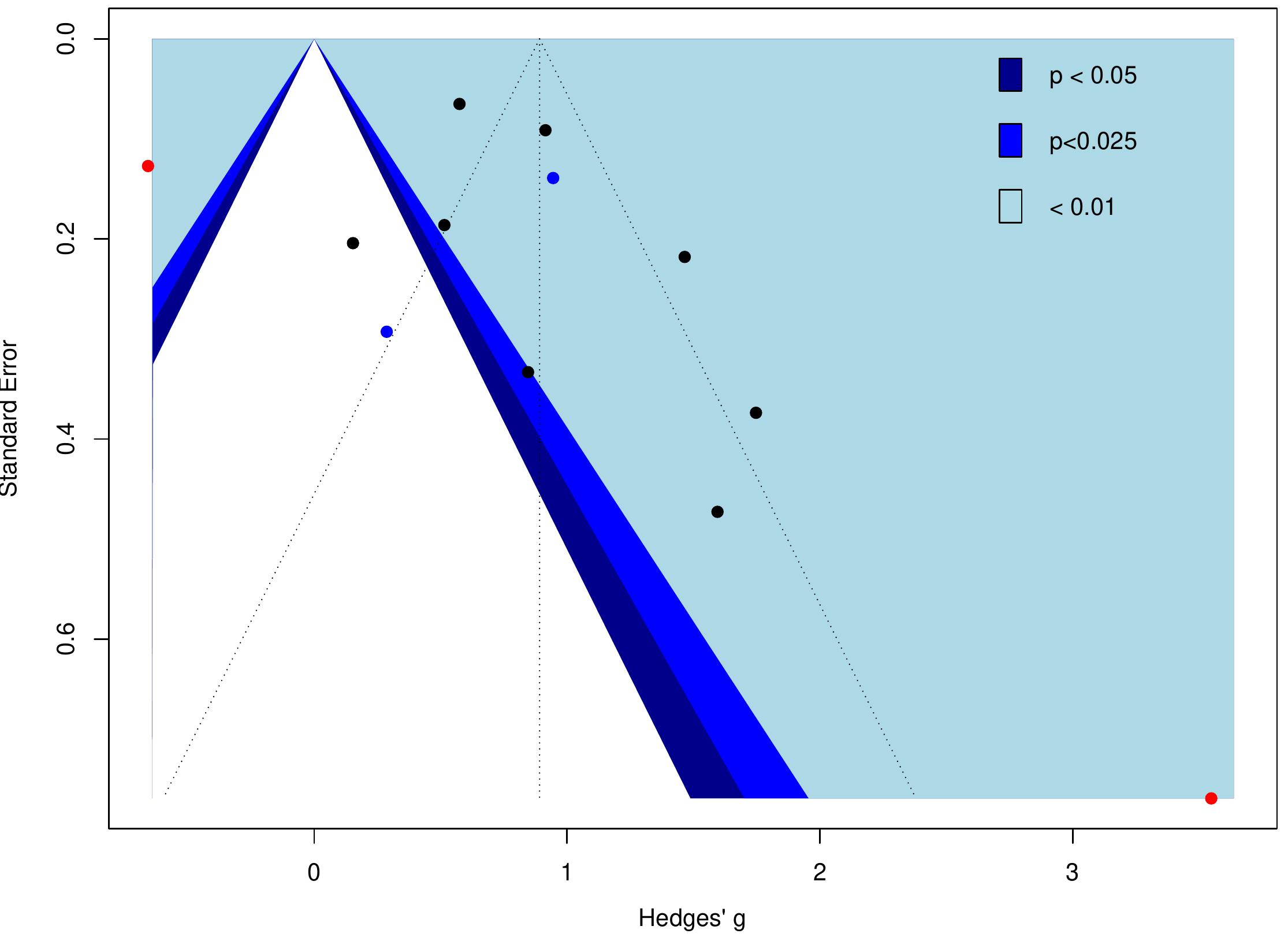}
    \caption{Funnel plot of the effect of training on phishing detection ability. Red points represent outliers and blue points unpublished works (dissertations).}~\label{figure:training_funnel}
\end{figure}

We also studied the publication bias using the Funnel plot (Figure \ref{figure:training_funnel}). Same as the analysis for \textit{gender}, we did not find a significant asymmetry in the plot using the Egger’s test (p=0.418).

\subsubsection{Warning and Highlighting}
Highlighting security clues (e.g., URL, sender email, etc.) or showing warning is another way of bringing users' attention to security cues. There are 12 studies that analyzed this (summarized in Table \ref{table:warning}). Eight of the studies found highlighting effective in improving users' susceptibility and the other four found it not useful. It is not easy to compare their results since there are lots of variables when we implement a highlighting or warning system. For example, researchers in \cite{williams2019persuasive,lin2011does} showed that just highlighting the domain name is not effective in reducing user click rate on phishing links. However, based on the results of \cite{volkamer2017user}, highlighting the domains and URLs combined with temporary disabling of links (to force users to read the warning) decreased users' susceptibility. We can conclude that highlighting can be useful to help people but it depends on how it is implemented.

\begin{table*}
\caption{Studies that analyzed the effect of highlighting and warning on phishing susceptibility (sorted based on their publication year). Green cells show significant improvements.}
\begin{threeparttable}
\resizebox{\textwidth}{!}{
\begin{tabular}{|c|r|c|c|c|c|c|}
\hline
Paper & Size & Vector & Highlighting Techniques& Gender dist. & Age Range & Statistics \\ \hline
\cellcolor{green!25}{Petelka et. al. '19\cite{petelka2019put}} &  701 & Email & Link-focused warning reduced phishing click rate & 56:42 & \begin{tabular}[l]{@{}c@{}}20-71\\ $\bar{x}=34.4$ \end{tabular}& Logistic Regression \\\hline
\cellcolor{red!25}{Williams and Polage '19\cite{williams2019persuasive}} & 178 & Email & Telling ``online risk news'' story to participants has no effect & 24:76 & \begin{tabular}[l]{@{}c@{}}$\geq$18\\(94\% in [18,24])\end{tabular} & ANOVA \\ \hline
\cellcolor{green!25}{Nicholson et. al. '17\cite{nicholson2017can}} & 281 & Email& Highlighting sender information made improvement & 58:40 & $\bar{x}=33.6$  &t-test \\\hline
\cellcolor{red!25}{Xiong et. al. '17\cite{xiong2017domain}} & 320 & Website& Domain highlighting has no effect & 42:58 & $\geq$18 & $\chi^2$ test \\\hline
\cellcolor{green!25}{Volkamer et. al. '17\cite{volkamer2017user}} & 86 & Email & \begin{tabular}[l]{@{}c@{}}Temporary URL disabling and highlighting link/domain \end{tabular} & 58:42 & \begin{tabular}[l]{@{}c@{}}17-60\\ ($\bar{x}=26.78$)\end{tabular} & Mann-Whitney U \\\hline
\cellcolor{green!25}{Yang et. al '17\cite{yang2017use}} & 63 & Email &  \begin{tabular}[l]{@{}c@{}}Suspicious URL warning using domain's traffic rank\tnote{a}\end{tabular} & 65:46 & 19-30 & $\chi^2$ test\\ \hline
\cellcolor{red!25}{Li et. al. '14\cite{li2014towards}} & 20 & Website & \begin{tabular}[l]{@{}c@{}}No difference between whitelist and  blacklist based toolbars\end{tabular} & 80:20 & 20-50 & t-test \\ \hline
\cellcolor{green!25}{Kirlappos and Sasse '11\cite{kirlappos2011security}} & 36 & Website & URL warning based on their popularity\tnote{b} & 47:53 & $\geq$18 ($\bar{x}=24$) & $\chi^2$ test \\\hline
\cellcolor{red!25}{Lin et. al. '10\cite{lin2011does}} & 22 & Website & Domain highlighting is not effective as a sole method & 77:23 & \begin{tabular}[l]{@{}c@{}}19-41\\ ($\bar{x}=27.7$, $\sigma=6.2$)\end{tabular} & ANOVA\\\hline
\cellcolor{green!25}{Egelman et. al. '08\cite{egelman2008you}} & 70 & Website & Active browser warning improved &  N/A & $\bar{x}=28$, $\sigma=10.58$ & Fisher's Exact Test\\\hline
\cellcolor{green!25}{Wu et. al. '06\cite{wu2006security}} & 30 & Website & Blocking warning box improved\tnote{c} & 53:47 & 18-50 ($\bar{x}=27$) & t-test \\\hline
\cellcolor{green!25}{Wu et. al. '06\cite{wu2006web}} & 21 & Website & \begin{tabular}[l]{@{}c@{}}Browser's auto-complete plugin that warns users in case of\\URL mismatch between saved credentials and opened URL\end{tabular} & 52:48 &N/A &  t-test\\\hline
\end{tabular}}
\begin{tablenotes}
\item[a] \textit{Authors did not mention the source of the domain rankings}
\item[b] \textit{Artificially generated links for the study purpose}
\item[c] \textit{Shows warning message to participants before opening a fraudulent website}
\end{tablenotes}
\end{threeparttable}
\label{table:warning}
\end{table*}

\begin{figure}
    \centering
    \includegraphics[width=\columnwidth]{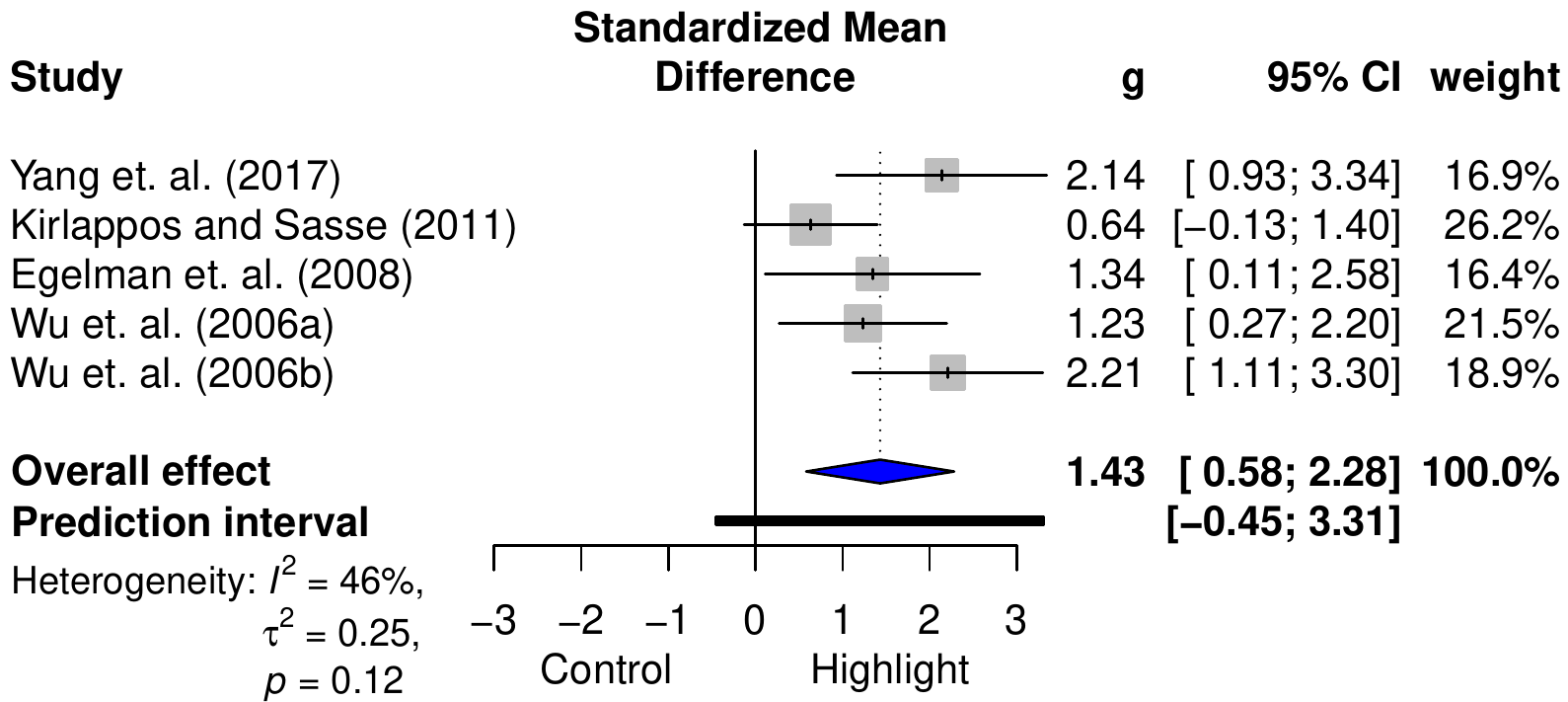}
    \caption{Forest plot of the effect size of \textit{highlighting} on phishing detection ability}~\label{fig:highlight_forest}
\end{figure}

For the meta-analysis, effect sizes could be calculated for only six (out of 12) of the studies that analyzed the warning effects because of the aforementioned issues. Figure \ref{fig:highlight_forest} shows the forest plot of the Random Effect Model. One study detected as influential on heterogeneity \cite{nicholson2017can}. This left us with four studies on website and only one on email \cite{yang2017use}. Based on the results obtained by combining the five studies, the improvement made on users' performance is significant ($t=4.68, p<0.01$). As reported in Table \ref{table:warning}, each work uses a different highlighting technique. Therefore, we cannot make a general conclusion that highlighting is useful in all cases. 

\section{Related Works}
Most of the survey papers on phishing focus on summarizing {\em automatic} email and website detection techniques, e.g.,  \cite{aassal20,sahoo2017malicious, dou17KK,mohammadTM15,almomaniGAMA13}. Despite considerable research on studying users' susceptibility to phishing attacks, there are only two surveys on these studies \cite{das2019,das2019all}. 

Authors in \cite{das2019} summarized 63 user study papers based on sample size, variables studied in the papers, types of participants (student or employees), environment (role-playing or real world) and the attack vector (email or website). They also selected eight of these studies based on certain criteria and summarized their findings. \hl{Comparing the attack vector of the studies, they also found more studies on phishing email than the website. Although they did a good job in providing a summary of the research on the \textit{phishing user study} area, their work cannot be compared with our analysis here. Their goal was to do systematic review and not quantitative comparison of the studies based on their findings.}

Researchers in \cite{das2019all} reviewed 51 user studies published in ACM Digital Library. They reviewed the papers using three attributes, 1) technical attributes of phishing attacks 2) individual attributes (age, gender, etc.) 3) spear phishing. \hl{Among the studies they reviewed, the also found a big ``striking lack of attention to reporting important information about methods and participants.'' They also found more studies on the technical attributes of phishing than on the individual aspects of users.} \hl{They reported how many papers studied each of these attributes and gave some example references.} \hl{Same as} \cite{das2019}{, their goal was also to do a literature review. So,} they did not analyze the findings of the papers, nor did they conduct any analysis of statistical significance.

\hl{To the best of our knowledge, there is no meta analysis on the phishing user studies}. In this work, we tried to fill this chasm in the literature by comparing 82 impactful papers and pointing out the missing information (e.g. subgroup sizes, the standard deviation in addition to mean) that could lead to a better analysis.

\section{Conclusions and Future Work}
In this study, we systematically reviewed user studies \hl{from four databases} on Internet users' phishing detection ability. We summarized the findings of 82 papers and reported their findings on the effect of five variables (\textit{age}, \textit{gender}, \textit{technological background}, \textit{training}, and \textit{warning}) on people's phishing/legitimate detection capability. We also did a meta-analysis using the Random Effect Model on 37 papers that we could calculate their effect sizes.

More than half of the papers that we reviewed failed to report their results comprehensively, so we could not calculate their effect size for meta-analysis. Problems include: missing subgroup sizes when they report the mean and standard deviation, missing mean and standard deviation for regression analysis, and missing data for  the significance test results (e.g.,  reporting p-value only). We recommend that researchers refer to \cite{wilson2017practical} as a guideline on how to report their results.  

\hl{Twenty four papers out of the 35} that compared the accuracy of males and females in detecting phishing and legitimate emails reported no significant difference. However, our meta-analysis results showed that males do perform better than females. For the \textit{age} variable, we also found older users performed better in distinguishing phishing emails/websites from legitimate instances.

Having more background knowledge about computer or Internet in general has been reported as a significant predictor of users susceptibility in the literature. But our meta-analysis showed that its effect is only marginally significant.  The meta-analysis of the \textit{training} effect on improving users' awareness showed that it significantly improves users detection. Further analysis showed that training users on websites has a greater effect than training on emails. 

The last variable that we studied was \textit{warning} techniques. Our results showed that the  warning also improves users' performance significantly. 


In this review of the literature, we observed that \textit{highlighting} is the least studied variable compared to other variables. More studies are required to better understand the \textit{highlighting}'s effectiveness on both email and website. Also, most of the studies that analyzed the relationship between \textit{background knowledge} as well as  \textit{age} on the phishing detection ability, did not report their results in a way that we could calculate effect sizes. More studies with comprehensive reports of results are necessary to better understand their relevance.

\ifCLASSOPTIONcaptionsoff
  \newpage
\fi



\bibliographystyle{IEEEtran}
\bibliography{IEEEabrv,bare_jrnl_compsoc.bib}
%

%








\end{document}